\documentclass[prl,%
twocolumn,%
tightenlines,nofootinbib,amsfonts,amsmath]{revtex4}
\usepackage{bm}\usepackage{epsfig}

\begin{document}

\title{La relativit\'e g\'en\'erale et la spirale infernale des
\'etoiles binaires compactes\\{\small{\rm [article paru dans ``Images de
la Physique 2005'', \'edit\'e par \'E. Falgarone {\em et al.},
\'Editions du C.N.R.S., p. 51 (2005)]}}}

\author{Luc Blanchet} \affiliation{${\mathcal{G}}{\mathbb{R}}
\varepsilon{\mathbb{C}}{\mathcal{O}}$, Institut d'Astrophysique de
Paris, 98$^{\text{bis}}$ boulevard Arago, 75014 Paris, France}

\date{\today}

\begin{abstract}
La th\'eorie relativiste de la gravitation, tr\`es bien v\'erifi\'ee par
les tests classiques dans le syst\`eme solaire et par le rayonnement
gravitationnel du pulsar binaire, est un des outils fondamentaux de
l'astrophysique. Elle permet le calcul de la forme de l'onde
gravitationnelle \'emise lors de la phase spiralante des binaires
d'\'etoiles \`a neutrons et de trous noirs. \`A partir d'une
approximation dite post newtonienne d\'evelopp\'ee \`a un ordre
\'elev\'e, la pr\'ediction de cette th\'eorie est utilis\'ee comme
``patron d'onde'' pour la recherche et l'analyse du signal dans le
r\'eseau de d\'etecteurs d'ondes gravitationnelles VIRGO/LIGO.
\end{abstract}

\maketitle

La relativit\'e g\'en\'erale est quelquefois consid\'er\'ee comme la
plus importante cr\'eation intellectuelle jamais r\'ealis\'ee par un
seul homme~: Albert Einstein. Elle a r\'evolutionn\'e notre vision de la
nature de l'espace et du temps, et de notre perception famili\`ere de la
force de la gravitation. Les physiciens ``relativistes'' admirent
l'extraordinaire coh\'erence math\'ematique -- et donc la beaut\'e -- de
ses \'equations. La relativit\'e g\'en\'erale est maintenant une vieille
dame, qui est n\'ee en 1915 apr\`es des ann\'ees de gestation laborieuse
remontant \`a la d\'ecouverte de la relativit\'e restreinte en 1905,
l'ann\'ee miraculeuse d'Einstein dont nous f\^etons le centenaire. Nous
allons voir que cette vieille dame est tr\`es en forme, car plus que
jamais la relativit\'e g\'en\'erale est consid\'er\'ee comme \textit{la}
th\'eorie de la gravitation.

Le ph\'enom\`ene familier de la gravitation poss\`ede en relativit\'e
g\'en\'erale l'interpr\'etation extraordinaire d'\^etre la manifestation
de la courbure de l'espace et du temps produite par la pr\'esence des
corps massifs. Cette description est une cons\'equence d'un principe
fondamental, appel\'e de nos jours le principe d'\'equivalence
d'Einstein, qui est la traduction en physique moderne du fait que tous
les corps sont acc\'el\'er\'es de la m\^eme fa\c{c}on dans un champ
gravitationnel. On dit parfois que la masse inerte $m_i$ des corps, qui
n'est rien d'autre que leur \'energie $E=m_i\,c^2$ en utilisant
l'\'equivalence masse-\'energie de la relativit\'e restreinte, est
toujours \'egale \`a leur masse gravitationnelle $m_g$, qui est
l'analogue de la charge \'electrique pour le champ gravitationnel. C'est
bien s\^ur Galil\'ee qui a fait remarquer l'importance de cette
``universalit\'e'' du mouvement de chute libre des corps (bien que son
exp\'erience fameuse du haut de la tour de Pise soit probablement
apocryphe), mais c'est Einstein qui a donn\'e \`a ce fait exp\'erimental
son statut d\'efinitif.

\section{LE TRIPTYQUE DES TESTS CLASSIQUES DANS LE SYST\`EME SOLAIRE} 

D\`es 1845, Le Verrier \`a l'observatoire de Paris (un an avant sa
d\'ecouverte de Neptune par le calcul \`a partir des perturbations
engendr\'ees sur Uranus), avait remarqu\'e que le demi grand-axe de
l'orbite de Mercure pr\'ecesse \`a chaque rotation avec un angle qui est
l\'eg\`erement en avance par rapport \`a la pr\'ediction th\'eorique
$\Delta_\mathrm{N}$. Son calcul de $\Delta_\mathrm{N}$, en th\'eorie de
Newton, \'etait fond\'e sur les perturbations induites par les autres
plan\`etes, principalement V\'enus qui est la plus proche de Mercure, et
Jupiter qui est la plus massive du syst\`eme solaire. L'avance anormale
du p\'erih\'elie \'etait rest\'ee inexpliqu\'ee et avait aliment\'e de
nombreuses sp\'eculations, parmi lesquelles l'existence d'une nouvelle
plan\`ete int\'erieure \`a l'orbite de Mercure (d\'enomm\'ee Vulcain par
Le Verrier), la pr\'esence possible d'un anneau de mati\`ere zodiacale
dans le plan de l'\'ecliptique, et m\^eme une modification de la loi
newtonienne en $1/r^2$. D\`es l'obtention des \'equations du champ
gravitationnel en novembre 1915, Einstein prouvera que les corrections
purement relativistes au mouvement d'une plan\`ete sur une ellipse
keplerienne impliquent une rotation suppl\'ementaire du grand axe de
l'ellipse donn\'ee par
$$\Delta_\mathrm{R} = \frac{6\pi GM_\odot}{c^2 a (1-e^2)}\,,$$ o\`u $a$
et $e$ sont le demi grand-axe et l'excentricit\'e de l'orbite, $M_\odot$
est la masse du Soleil, et $G$ et $c$ sont la constante de la
gravitation et la vitesse de la lumi\`ere. Num\'eriquement on trouve
43'' d'arc par si\`ecle, qui s'ajoutent donc \`a la pr\'ecession
newtonienne $\Delta_\mathrm{N}$ pour \^etre en parfait accord avec
l'observation~! C'est certainement ce succ\`es remarquable qui a
convaincu Einstein de la justesse de la th\'eorie naissante (c'\'etait
d'ailleurs \`a l'\'epoque la seule confrontation possible de la
th\'eorie \`a des observations r\'eelles).

Le deuxi\`eme ``test classique'', encore plus c\'el\`ebre, est celui de
l'angle de d\'eviation de la lumi\`ere en provenance d'une source lointaine
(un quasar dans les mesures r\'ecentes), par le champ de gravitation du
Soleil. Il est donn\'e en relativit\'e g\'en\'erale, dans le cas d'un rayon
rasant la surface du Soleil (rayon $R_\odot$), par
$$\alpha_\odot = \frac{4GM_\odot}{c^2 R_\odot}\,.$$ Cet angle vaut
\textit{deux} fois la valeur estim\'ee en th\'eorie de Newton, car en
effet si on consid\`ere la lumi\`ere comme faite de corpuscules de
vitesse $c$ (et de masse arbitraire, car la masse n'intervient pas), il
y a bien une d\'eviation de la lumi\`ere chez Newton~! En fait on peut
montrer que le facteur 4 dans l'expression de $\alpha_\odot$ se
d\'ecompose en ``2+2'', avec le premier 2 qui provient du principe
d'\'equivalence, $m_i=m_g$, qui est vrai en relativit\'e g\'en\'erale
comme en th\'eorie de Newton, et le second 2 qui est un effet
suppl\'ementaire d\^u \`a la courbure de l'espace en relativit\'e
g\'en\'erale. L'angle $\alpha_\odot$ vaut 1.75'' d'arc, et fut mesur\'e
lors d'une \'eclipse du Soleil par Eddington en 1919, qui put d'ores et
d\'ej\`a conclure que la th\'eorie de Newton \'etait exclue
exp\'erimentalement. (En cette ann\'ee du trait\'e de Versailles un
anglais mettait \`a mal la th\'eorie d'un autre anglais, et confirmait
exp\'erimentalement celle d'un allemand.)

L'effet Shapiro compl\`ete notre triptyque des tests classiques de la
relativit\'e g\'en\'erale dans le syst\`eme solaire. C'est un retard
d\^u au champ de gravitation dans les temps d'arriv\'ee de photons ayant
ras\'e la surface du Soleil. Non seulement la trajectoire de la
lumi\`ere est d\'evi\'ee de l'angle $\alpha_\odot$, mais les photons sur
leur trajectoire sont \textit{ralentis} par le champ du Soleil. L'effet
n'est pas du tout n\'egligeable, et il a \'et\'e calcul\'e et observ\'e
pour la premi\`ere fois par Shapiro en 1964. Son exp\'erience a
consist\'e \`a mesurer le temps d'aller-retour de photons radio \'emis
sur Terre vers Mercure, r\'efl\'echis sur le sol de Mercure et
renvoy\'es vers la Terre, lorsque la trajectoire des photons passe \`a
proximit\'e de la surface du Soleil. L'effet principal du ralentissement
de la lumi\`ere est donn\'e par
$$\Delta T =
\frac{4GM_\odot}{c^3}\log\left(\frac{4\,r_\oplus\,r_\otimes}{{R_\odot}^{\!2}}
\right)\,,$$ o\`u $r_\oplus$ et $r_\otimes$ sont les distances de la
Terre et de Mercure au Soleil. Contrairement aux autres tests, l'effet
Shapiro ne date pas de l'enfance de la relativit\'e g\'en\'erale.
Curieusement, Einstein n'a jamais pens\'e \`a calculer cet effet. Ayant
obtenu la trajectoire des photons au voisinage du Soleil et leur angle
de d\'eviation $\alpha_\odot$, il n'a apparemment jamais cherch\'e \`a
conna\^itre le mouvement ``horaire'' des photons sur leur trajectoire,
ce qui lui aurait donn\'e leur retard gravitationnel $\Delta T$ -- un
nouvel effet tout \`a fait int\'eressant.

La relativit\'e g\'en\'erale est maintenant v\'erifi\'ee dans le
syst\`eme solaire \`a mieux que 1/1000$^{\text{\`eme}}$ pr\`es. Des
mesures tr\`es pr\'ecises d'astrom\'etrie, telles que celles du futur
satellite GAIA qui sera lanc\'e par l'agence spatiale europ\'eenne,
devraient encore am\'eliorer la pr\'ecision sur la d\'eviation de la
lumi\`ere. Des th\'eories alternatives de la gravitation, comme par
exemple la th\'eorie de Brans et Dicke o\`u l'on rajoute un champ
scalaire au champ gravitationnel de la relativit\'e g\'en\'erale, et qui
fut une th\'eorie fameuse en son temps (1961), ont \'et\'e \'elimin\'ees
par ces observations. Cependant, dans le syst\`eme solaire, les vitesses
des corps sont tr\`es petites par rapport \`a la vitesse de la
lumi\`ere, $v\lesssim 10^{-4}\,c$, et le champ de gravitation est
faible, car le potentiel newtonien $U$ est tout petit en unit\'es
relativistes, $U\lesssim 10^{-6}\,c^2$. Les tests classiques n'ont donc
v\'erifi\'e qu'un r\'egime assez restreint de la th\'eorie, celui de sa
limite ``quasi-newtonienne''.

\section{LA RELATIVIT\'E G\'EN\'ERALE, UN OUTIL POUR
L'ASTROPHYSIQUE} 

Apr\`es son enfance brillante, la vieille dame a connu une adolescence
difficile. Elle fut longtemps consid\'er\'ee comme un ``paradis pour le
th\'eoricien'', mais un ``d\'esert pour l'exp\'erimentateur''. En fait,
elle est rest\'ee \`a l'\'ecart du courant principal de la physique,
domin\'e par la m\'ecanique quantique et la th\'eorie quantique des
champs, jusqu'au d\'ebut des ann\'ees 1960 (notre th\'eorie est alors
quadrag\'enaire), \'epoque \`a laquelle elle a subi un renouveau et un
essor remarquables.

Du point de vue th\'eorique, cette \'epoque a vu l'\'elucidation du
concept de trou noir, et la magnifique d\'ecouverte par Kerr du trou
noir en rotation (1963). Le trou noir de Schwarzschild, sans rotation,
date des premiers mois de la relativit\'e g\'en\'erale, mais \`a
l'\'epoque on consid\'erait cette solution comme valable uniquement \`a
l'ext\'erieur d'une \'etoile, et ce n'est que dans les ann\'ees 60 que
l'on analysera les propri\'et\'es du trou noir au voisinage de
l'horizon, pour comprendre que ces objets peuvent r\'eellement exister
dans la nature. De m\^eme pour le rayonnement gravitationnel, dont on a
vraiment compris les caract\'eristiques pendant cette p\'eriode;
auparavant une controverse faisait rage sur l'existence r\'eelle des
ondes gravitationnelles~!

On peut dire que les exp\'eriences modernes de gravitation ont
commenc\'e avec la v\'erification pr\'ecise en laboratoire du d\'ecalage
gravitationnel vers le rouge ou effet Einstein, par Pound et Rebka en
1960. Souvent consid\'er\'ee comme le 4$^{\text{\`eme}}$ test classique
de la th\'eorie, cette v\'erification est en fait un test du principe
d'\'equivalence qui est donc plus g\'en\'eral. \`A la m\^eme \'epoque on
tentait la d\'etection du rayonnement gravitationnel \`a l'aide d'un
cylindre m\'etallique r\'esonnant appel\'e maintenant barre de Weber.

La relativit\'e g\'en\'erale \'emerge alors enfin en tant que th\'eorie
\textit{physique}, qui fait des pr\'edictions et voit ses pr\'edictions
r\'ealis\'ees. La d\'ecouverte en 1974 du pulsar binaire PSR~1913+16, et
la preuve exp\'erimentale de l'existence du rayonnement gravitationnel
tel qu'il est pr\'evu par la relativit\'e g\'en\'erale, illustre
merveilleusement la capacit\'e de pr\'ediction de notre th\'eorie (voir
la section suivante).

Aujourd'hui la relativit\'e g\'en\'erale est un ``outil'' permettant
d'explorer l'existence et de comprendre les observations de nouveaux
objets ou de nouveaux ph\'enom\`enes en astrophysique. Par exemple les
propri\'et\'es particuli\`eres du trou noir de Kerr sont utilis\'ees par
les astrophysiciens travaillant sur les objets compacts et les disques
d'accr\'etion autour de trous noirs. La relativit\'e g\'en\'erale va
probablement permettre d'ouvrir une nouvelle ``fen\^etre'' en
astronomie, celle des ondes gravitationnelles, car ce rayonnement a des
propri\'et\'es sp\'ecifiques tr\`es diff\'erentes des ondes
\'electromagn\'etiques.

Il faut pourtant garder \`a l'esprit que le domaine o\`u s'exerce la
relativit\'e g\'en\'erale est le \textit{macrocosme}. Cette th\'eorie
n'incorpore pas les lois de la m\'ecanique quantique, et il est probable
qu'elle doive \^etre consid\'er\'ee comme une th\'eorie ``effective''
valable uniquement \`a grande \'echelle. Assez \'etrangement, la force
gravitationnelle n'a pu \^etre test\'ee en laboratoire que jusqu'\`a une
\'echelle de l'ordre du millim\`etre. \`A une \'echelle microscopique,
inf\'erieure ou tr\`es inf\'erieure au millim\`etre, on ne conna\^it
exp\'erimentalement rien de la loi gravitationnelle et il est
vraisemblable que la relativit\'e g\'en\'erale \textit{stricto sensu} ne
s'applique plus.

\section{LE PULSAR BINAIRE PSR~1913+16} 

L'ann\'ee 1974 fut faste pour les ``relativistes'' avec la d\'ecouverte
par Hulse et Taylor d'un syst\`eme extr\^emement int\'eressant~: le
pulsar binaire PSR~1913+16, qui valut \`a ses d\'ecouvreurs le prix
Nobel en 1993. C'est un pulsar, c'est-\`a-dire une \'etoile \`a neutrons
en rotation rapide sur elle-m\^eme (avec une p\'eriode de
$56\,\mathrm{ms}$), qui envoie \`a chaque rotation, tel un phare, du
rayonnement \'electromagn\'etique radio en direction de la Terre.
L'analyse des instants d'arriv\'ee des pulses radio montre (gr\^ace \`a
leur d\'ecalage Doppler) que PSR~1913+16 est en orbite autour d'une
\'etoile compagnon, probablement une autre \'etoile \`a neutrons.
L'orbite est une ellipse quasi-keplerienne de p\'eriode orbitale
$P\simeq 7^\mathrm{h}40^\mathrm{mn}$, d'excentricit\'e $e\simeq 0.617$
et de demi grand-axe $a\simeq 10^6\,\mathrm{km}$. Les masses du pulsar
et de son compagnon ($m_p$ et $m_c$) sont toutes deux environ \'egales
\`a $1.4\,M_\odot$ (qui est la masse des \'etoiles \`a neutrons).
PSR~1913+16 est un syst\`eme passionnant car les effets relativistes
jouent un r\^ole important dans sa dynamique. Par exemple, la
pr\'ecession relativiste $\Delta_\mathrm{R}$ du p\'eriastre de l'orbite
est de l'ordre de 4 degr\'es par an, \`a comparer avec les 43'' arc par
si\`ecle du p\'erih\'elie de Mercure.

Le syst\`eme double form\'e par le pulsar et son compagnon \'emet du
rayonnement gravitationnel, ce qui se traduit par une perte d'\'energie
orbitale, et donc par le rapprochement des deux \'etoiles l'une de
l'autre, et une lente d\'erive de la p\'eriode orbitale du mouvement
($\dot{P}<0$). On sait qu'en premi\`ere approximation le rayonnement
gravitationnel est quadrupolaire et le flux du rayonnement est donn\'e
par la formule dite du quadrupole d'Einstein. Si l'on applique cette
formule \`a un syst\`eme de deux masses ponctuelles en mouvement sur une
ellipse keplerienne on trouve un r\'esultat d\^u \`a Peters et Mathews
(1963),
$$\dot{P}=-\frac{192\pi}{5c^5} \left(\frac{2\pi
G}{P}\right)^{5/3}\!\!\!\!
\frac{m_p\,m_c}{(m_p+m_c)^{1/3}}\frac{1+\frac{73}{24}e^2+\frac{37}{96}
e^4}{(1-e^2)^{7/2}}\,.$$ Dans le cas du pulsar binaire cette formule
donne $\dot{P}=-2.4~10^{-12}\,\mathrm{s/s}$, qui repr\'esente donc la
d\'ecroissance de la p\'eriode orbitale mesur\'ee en secondes \`a chaque
seconde. Cette pr\'ediction purement th\'eorique est en excellent accord
(\`a mieux que $0.5\%$ pr\`es) avec les observations effectu\'ees par
Taylor et ses collaborateurs. C'est une v\'erification remarquable,
l'une des confirmations les plus importantes de la relativit\'e
g\'en\'erale, et l'une des mesures les plus pr\'ecises effectu\'ees en
astronomie. Mais pour nous elle repr\'esente une validation
observationnelle de l'ordre newtonien sur lequel est fond\'e nos
investigations post newtoniennes pour les binaires compactes spiralantes
(voir la derni\`ere section)~!

\section{LA D\'ETECTION INTERF\'EROM\`ETRIQUE DU RAYONNEMENT
GRAVITATIONNEL}

Une onde gravitationnelle est engendr\'ee par le mouvement
acc\'el\'er\'e des corps massifs. Dans notre th\'eorie, le champ de
gravitation est repr\'esent\'e par la m\'etrique de l'espace-temps, et
l'on peut montrer que les composantes de la m\'etrique ob\'eissent en
premi\`ere approximation, dans un syst\`eme de coordonn\'ees
particulier, \`a une \'equation de d'Alembert ou \'equation des ondes.
On a donc affaire \`a une onde gravitationnelle, qui doit \^etre vue
comme une perturbation de la surface de l'espace-temps se propageant \`a
la vitesse de la lumi\`ere. L'action de l'onde gravitationnelle sur la
mati\`ere se traduit par des d\'eformations analogues \`a celles
produites par un champ de mar\'ee. La variation relative de la taille
$L$ d'un d\'etecteur au passage de l'onde est donn\'ee par
$$\frac{\delta L}{L}\simeq\frac{h}{2}\,,$$ o\`u $h$ repr\'esente l'amplitude
de l'onde gravitationnelle, c'est-\`a-dire la modification de la m\'etrique de
l'espace-temps par rapport \`a la m\'etrique ``plate'' de Minkowski en
l'absence du champ gravitationnel.

De nouvelles exp\'eriences vont tenter pour la premi\`ere fois de
d\'etecter le rayonnement gravitationnel produit par des sources
cosmiques. Ces exp\'eriences sont fond\'ees sur l'interf\'erom\`etrie
\`a laser, et sont constitu\'ees de gigantesques interf\'erom\`etres de
Michelson avec cavit\'es Fabry-Perot. Elles ambitionnent de former un
r\'eseau international, comprenant des interf\'erom\`etres de grande
taille, LIGO aux \'Etats-Unis dont les bras ont une longueur de
$4\,\mathrm{km}$, et VIRGO qui est construit pr\`es de Pise avec des
bras de $3\,\mathrm{km}$ dans le cadre d'une collaboration
franco-italienne (voir la figure 1). Le r\'eseau comprend aussi des
d\'etecteurs de taille plus modeste, GEO \`a Hanovre et TAMA au Japon.

Le grand int\'er\^et de l'interf\'erom\'etrie \`a laser pour la
d\'etection des ondes gravitationnelles, est la bande de fr\'equence
tr\`es large du d\'etecteur, typiquement de $\sim 10\,\mathrm{Hz}$ \`a
$1000\,\mathrm{Hz}$ pour VIRGO. Les barres de Weber en revanche ne
peuvent d\'etecter qu'au voisinage de la fr\'equence de r\'esonance de
la barre.
\begin{figure}
\begin{center}
\epsfig{file=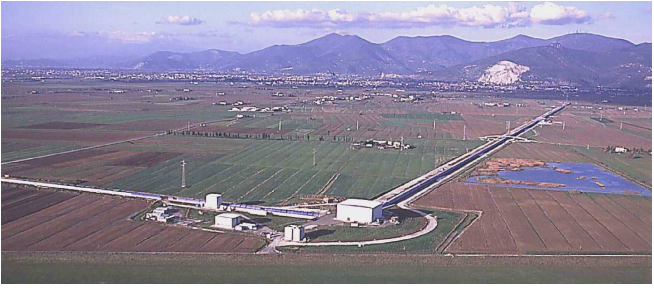,width=3.4in}
\caption{Vue du d\'etecteur d'ondes gravitationnelles VIRGO
\`a Cascina, pr\`es de Pise.}
\end{center}
\end{figure}

L'amplitude attendue pour une onde gravitationnelle en provenance de
syst\`emes binaires \`a une distance de 100 Mpc est $h\sim 10^{-23}$, ce
qui, d'apr\`es l'estimation pr\'ec\'edente sur la variation de longueur
du d\'etecteur, donne $\delta L\sim 10^{-20}\,\mathrm{m}$ dans le cas de
VIRGO soit $10^{-5}\,\mathrm{fermi}$~! Comment est-il possible de
mesurer un tel d\'eplacement~? La r\'eponse est qu'en r\'ealit\'e on
mesure le d\'eplacement \textit{collectif} de $N$ atomes des miroirs en
entr\'ee et en bout de bras de l'interf\'erom\`etre. La mesure contient
donc un effet de moyenne sur les atomes ce qui permet de gagner un
facteur $\sqrt{N}$. Avec $N\sim 10^{18}$ ce qui correspond \`a une
couche atomique en surface du miroir on constate que la mesure effective
\`a r\'ealiser est beaucoup plus raisonnable, $\delta
L_\mathrm{eff}=\sqrt{N}\,\delta L\sim 10^{-11}\,\mathrm{m}$.

Il existe de nombreuses sources astrophysiques potentielles dont le
rayonnement gravitationnel pourrait \^etre d\'etect\'e par VIRGO et
LIGO. Les supernov\ae, qui sont des explosions d'\'etoiles massives en
fin de vie lorsqu'elles ont \'epuis\'e tout leur ``combustible''
nucl\'eaire, ont longtemps \'et\'e consid\'er\'ees comme des sources
d'ondes gravitationnelles int\'eressantes, mais on sait maintenant
qu'elles engendrent en fait peu de rayonnement. En effet l'effondrement
des couches internes de la supernova, qui devrait \^etre responsable de
la production du rayonnement gravitationnel, est essentiellement
sph\'erique, et d'apr\`es un th\'eor\`eme fameux de relativit\'e
g\'en\'erale, le champ ext\'erieur \`a une distribution sph\'erique de
mati\`ere est donn\'e par la solution de Schwarzschild qui est
\textit{statique} -- il n'y a donc pas de rayonnement \'emis. Beaucoup
plus int\'eressants pour VIRGO et LIGO sont les syst\`emes binaires, car
leur dynamique est fortement asym\'etrique et ils engendrent beaucoup de
rayonnement gravitationnel.

\section{SPIRALE ET MORT DES SYST\`EMES BINAIRES D'\'ETOILES
COMPACTES} 

Une binaire compacte spiralante est ce que deviendra le pulsar binaire
PSR~1913+16 dans quelques centaines de millions d'ann\'ees, lorsqu'il
finira par fusionner avec son compagnon. Pendant toute sa vie il aura
\'emis son \'energie de liaison gravitationnelle sous forme de
rayonnement gravitationnel, jusqu'\`a ce que les deux \'etoiles \`a
neutrons tombent l'une sur l'autre. (Rappelons qu'une \'etoile \`a
neutrons est un astre compact, form\'e essentiellement de neutrons avec
une densit\'e comparable \`a celle de la mati\`ere nucl\'eaire, et dont
la taille est \`a peu pr\`es celle de l'agglom\'eration parisienne pour
une masse de $\sim 1.4\,M_\odot$.)

Dans les derniers instants avant la fusion finale, les deux objets
compacts (\'etoiles \`a neutrons ou trous noirs) d\'ecrivent une orbite
rapproch\'ee qui a la forme d'une spirale circulaire rentrante \`a cause
de la perte d'\'energie li\'ee \`a l'\'emission du rayonnement
gravitationnel. C'est ce rayonnement que l'on observera sur Terre o\`u
il d\'eformera l'espace-temps avec une amplitude relative de l'ordre de
$10^{-23}$ (voir la figure 2). Au cours de la phase spiralante, la
distance entre les deux \'etoiles diminue au cours du temps, et la
fr\'equence orbitale du mouvement, $\omega=2\pi/P$ o\`u $P$ est la
p\'eriode, augmente. On peut montrer que l'\'evolution de l'orbite est
adiabatique, dans le sens o\`u le changement relatif de fr\'equence
pendant une p\'eriode correspondante reste faible,
$\dot{\omega}/\omega^2 \lesssim 0.1$. Nous allons voir comment cette
propri\'et\'e d'adiabaticit\'e permet de d\'efinir un sch\'ema
d'approximation tr\`es puissant en relativit\'e g\'en\'erale, capable de
d\'ecrire la spirale avec grande pr\'ecision.
\begin{figure}
\begin{center}
\epsfig{file=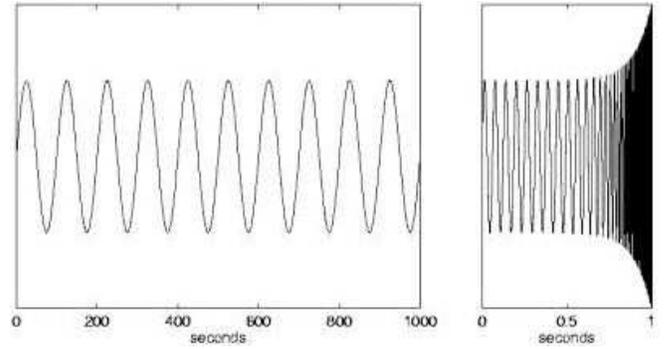,width=3.4in}
\caption{Onde gravitationnelle $h(t)$ \'emise par une binaire compacte
spiralante. La fr\'equence et l'amplitude de l'onde augmentent
adiabatiquement au cours du temps.}
\end{center}
\end{figure}

\`A la fin de la phase spiralante, le syst\`eme binaire atteint ce qu'on
appelle la derni\`ere orbite circulaire (dite aussi ``innermost circular
orbit'' ou ICO), \`a partir de laquelle l'\'evolution de l'orbite cesse
d'\^etre adiabatique. Les deux corps plongent alors l'un sur l'autre et
fusionnent tr\`es rapidement pour former un trou noir unique. \`A cause
de la dynamique violente qui conduit \`a la formation de ce trou noir,
celui-ci est initialement tr\`es d\'eform\'e et soumis \`a d'importantes
vibrations, mais il finira par atteindre, avec l'\'emission de ses modes
de vibration intrins\`eques en ondes gravitationnelles, un r\'egime
stationnaire d\'ecrit par la solution de Kerr pour le trou noir en
rotation.

Les binaires compactes spiralantes sont des syst\`emes parmi les plus
relativistes que l'on puisse imaginer, \`a la fois du point de vue de la
relativit\'e restreinte, car la vitesse orbitale atteint $\sim 0.5\,c$
au moment du passage \`a la derni\`ere orbite circulaire, et de la
relativit\'e g\'en\'erale, car les masses en jeu sont importantes,
$1.4\,M_\odot$ pour les \'etoiles \`a neutrons et peut-\^etre jusqu'\`a
$20\,M_\odot$ pour des trous noirs, donc les champs gravitationnels sont
intenses. Des milliers de cycles orbitaux sont parcourus en quelques
secondes avant la fusion finale, dans ce qu'on peut d\'ecrire de
fa\c{c}on imag\'ee comme une ``spirale infernale'', constituant les
derniers spasmes de l'agonie du syst\`eme binaire. La spirale et la mort
des binaires compactes repr\'esente donc un \'ev\'enement tout \`a fait
impressionnant du point de vue de l'astrophysique. C'est principalement
pendant la phase spiral\'ee pr\'ec\'edant imm\'ediatement la fusion que
l'onde gravitationnelle qui sera d\'etect\'ee par VIRGO et LIGO est
produite.

De tels \'ev\'enements cataclysmiques de fusion d'objets compacts se
produisent dans l'univers. \`A partir des syst\`emes binaires
d'\'etoiles \`a neutrons connus dans notre Galaxie (le pulsar binaire
PSR~1913+16 en est l'exemple le plus c\'el\`ebre), on peut en d\'eduire
que quelques fusions d'\'etoiles \`a neutrons devraient survenir par an
dans un volume de environ $100\,\mathrm{Mpc}$ de rayon centr\'e sur
notre Galaxie. Un tel volume contient des milliers de galaxies, dont
toutes celles de l'amas de la Vierge (qui a donn\'e son nom \`a VIRGO)
situ\'e au centre du super-amas de galaxies dans lequel nous vivons. \`A
cette distance le signal gravitationnel sera assez puissant pour \^etre
observ\'e par le r\'eseau actuel des d\'etecteurs ou par une
g\'en\'eration de d\'etecteurs de sensibilit\'e am\'elior\'ee. Il
rentrera dans leur bande de fr\'equence quelques minutes avant la
fusion, lorsque la fr\'equence du signal gravitationnel atteindra $f\sim
10\,\mathrm{Hz}$ (d'apr\`es les propri\'et\'es des ondes
gravitationnelles la fr\'equence de l'harmonique principale du signal
est double de la fr\'equence orbitale du mouvement, $f=\omega/\pi=2/P$).
VIRGO, qui a de tr\`es bonnes performances \`a basse fr\'equence gr\^ace
\`a ses tours d'isolation du bruit sismique terrestre, pourra notamment
capter tr\`es t\^ot le signal des binaires spiralantes d'\'etoiles \`a
neutrons, et augmenter ainsi le rapport signal-sur-bruit par une longue
int\'egration du signal \`a partir de la basse fr\'equence.

L'existence des binaires spiralantes de trous noirs est plus incertaine,
car malheureusement on ne conna\^it pas de syst\`emes de deux trous
noirs dans notre Galaxie. N\'eanmoins, gr\^ace \`a la simulation
num\'erique des phases successives d'\'evolution des syst\`emes
binaires, on peut estimer que le taux de fusion de deux trous noirs
pourrait \^etre comparable ou sup\'erieur \`a celui des \'etoiles \`a
neutrons. Les astrophysiciens pensent souvent que la premi\`ere
d\'etection du rayonnement gravitationnel par VIRGO et LIGO sera celle
d'une binaire spiralante de trous noirs.

\section{LE PROBL\`EME DES DEUX CORPS EN RELATIVIT\'E
G\'EN\'ERALE}

Pour le th\'eoricien ``relativiste'' l'int\'er\^et principal des
binaires compactes spiralantes r\'eside dans le fait que l'onde
gravitationnelle qu'elles \'emettent est \textit{calculable} avec grande
pr\'ecision, ind\'ependamment des d\'etails de la structure interne des
\'etoiles compactes, et de la pr\'esence possible d'un environnement
astrophysique ``sale''. Par exemple les effets non gravitationnels, qui
compliquent habituellement l'astrophysique des syst\`emes binaires
(champs magn\'etiques, pr\'esence d'un milieu interstellaire,
\textit{etc.}), sont n\'egligeables. La dynamique des binaires
spiralantes est domin\'ee par les effets gravitationnels orbitaux (les
effets de mar\'ees jouent tr\`es peu de r\^ole dans le cas de corps
compacts). On peut donc mod\'eliser le syst\`eme par deux particules
ponctuelles, sans structure interne, caract\'eris\'ees uniquement par
leurs masses $m_1$ et $m_2$, et aussi \'eventuellement par leurs spins.

Ainsi le probl\`eme th\'eorique des binaires spiralantes est un pur
probl\`eme de m\'ecanique c\'eleste~: le probl\`eme des deux corps
(consid\'er\'es comme ponctuels) en relativit\'e g\'en\'erale. On sait
qu'en th\'eorie de Newton le probl\`eme des 2 corps est
``int\'egrable'', mais qu'\`a partir de 3 corps les \'equations du
mouvement ne peuvent pas \^etre r\'esolues dans le cas g\'en\'eral. En
relativit\'e g\'en\'erale m\^eme pour 2 corps on ne peut pas \'ecrire de
fa\c{c}on exacte les \'equations du mouvement et encore moins les
r\'esoudre~! Le probl\`eme \`a un corps en revanche admet une solution
exacte qui est donn\'ee par la m\'etrique du trou noir de Schwarzschild.

N'admettant pas de solution exacte le probl\`eme relativiste des deux
corps doit \^etre trait\'e par des m\'ethodes d'approximations. Ce n'est
pas un drame car pratiquement tous les grands succ\`es de la
relativit\'e g\'en\'erale dans sa confrontation avec l'exp\'erience et
l'observation ont \'et\'es obtenus gr\^ace \`a de telles m\'ethodes.

Notre sch\'ema d'approximation fait intervenir ce que nous avons dit \`a
propos du spiralement adiabatique des binaires compactes. Il se trouve
en effet que le petit param\`etre adiabatique est en fait petit dans le
sens \textit{post newtonien}, car on a $\dot{\omega}/\omega^2 =
\mathcal{O}\left[(v/c)^5\right]$, o\`u $v$ est la vitesse orbitale des
corps et $c$ la vitesse de la lumi\`ere. L'approximation post
newtonienne consiste \`a d\'evelopper la relativit\'e g\'en\'erale
autour de la th\'eorie de Newton, sous la forme d'un d\'eveloppement en
puissances de $v/c$ lorsque $v/c\rightarrow 0$, ce qui peut \^etre vu
plus formellement comme un d\'eveloppement quand $c\rightarrow +\infty$.
L'ordre d'approximation $(v/c)^5$ correspond au premier effet de ce
qu'on appelle la \textit{r\'eaction de rayonnement}, c'est-\`a-dire
l'influence de l'\'emission du rayonnement gravitationnel sur le
mouvement du syst\`eme binaire, qui se traduit par un petit effet de
``freinage'' des corps sur leur orbite, et donc par une d\'ecroissance
de la p\'eriode orbitale $P$.

L'approximation post newtonienne est la seule technique connue qui
permet de d\'ecrire la phase spiralante des binaires compactes (pendant
laquelle on a $v/c\lesssim 0.5$), et elle est valable jusqu'\`a la
derni\`ere orbite circulaire. Pass\'ee cette orbite le d\'eveloppement
post newtonien devrait en principe \^etre remplac\'e par un calcul
d'int\'egration \textit{num\'erique} des \'equations d'Einstein. Un tel
calcul est indispensable pour d\'ecrire en d\'etail le m\'ecanisme de
fusion des deux horizons des trous noirs, et obtenir la forme d'onde
gravitationnelle produite lors de cette phase. Malheureusement la
relativit\'e num\'erique n'a pas encore r\'eussi \`a r\'esoudre ce
probl\`eme extr\^emement difficile, bien qu'il ait \'et\'e l'objet de ce
qu'on a appel\'e le ``binary black hole Grand challenge'', qui a
mobilis\'e de nombreux instituts am\'ericains mais n'a pas apport\'e les
r\'esultats escompt\'es. Il se trouve que vouloir calculer
num\'eriquement la fusion de deux trous noirs en utilisant la ``force
brute'' d'un ordinateur n'est pas r\'ealisable actuellement, malgr\'e
certaines perc\'ees remarquables ces derni\`eres ann\'ees. Heureusement,
dans le cas d'\'etoiles \`a neutrons ou de trous noirs peu massifs, la
plus grande partie du rapport signal-sur-bruit dans VIRGO et LIGO
proviendra de la phase spiralante pr\'ec\'edant la fusion, qui est
tr\`es bien d\'ecrite par la th\'eorie post newtonienne.

\section{PATRONS D'ONDES GRAVITATIONNELLES POUR VIRGO ET LIGO} 

Le d\'eveloppement post newtonien va s'av\'erer l'outil id\'eal pour le
calcul de la radiation gravitationnelle d'une binaire compacte
spiralante. Et comme l'approximation post newtonienne va devoir \^etre
d\'evelopp\'ee jusqu'\`a un ordre tr\`es \'elev\'e ce probl\`eme va
devenir un vrai paradis pour le th\'eoricien~! Des \'etudes d'analyse du
signal dans les d\'etecteurs VIRGO et LIGO ont en effet montr\'e qu'une
pr\'ediction tr\`es pr\'ecise de la relativit\'e g\'en\'erale est
n\'ecessaire pour tirer parti de toute l'information potentielle
contenue dans le signal des binaires spiralantes. Pour d\'etecter le
signal gravitationnel (et l'analyser ult\'erieurement) on utilise la
pr\'ediction th\'eorique, que l'on appelle pour l'occasion le ``patron''
d'onde gravitationnelle, et on effectue sa corr\'elation avec le signal
de sortie du d\'etecteur. Si le patron est une copie fid\`ele du signal
r\'eel (c'est-\`a-dire si la pr\'ediction de la relativit\'e
g\'en\'erale est correcte), alors la corr\'elation est importante, et
l'on aura d\'etect\'e une onde gravitationnelle.

Il a \'et\'e montr\'e que les patrons d'ondes doivent prendre en compte
toutes les corrections relativistes dans le champ d'ondes
gravitationnelles jusqu'\`a la troisi\`eme approximation post
newtonienne, qui correspond \`a une pr\'ecision relative incluant tous
les termes jusqu'\`a l'ordre $(v/c)^6$ par rapport \`a la formule du
quadrupole d'Einstein pour le rayonnement gravitationnel. Dans le jargon
cette approximation s'appelle 3PN, et plus g\'en\'eralement les termes
post newtoniens $\sim (v/c)^{2n}$ sont dits d'ordre $n$PN.

Pour mieux comprendre ce que signifie la pr\'ecision 3PN, rappelons-nous
que la formule du quadrupole d'Einstein d\'ecrit l'\'emission des ondes
gravitationnelles \`a l'ordre dominant (quadrupolaire en relativit\'e
g\'en\'erale), qui est newtonien dans le sens o\`u \`a cet ordre
d'approximation le quadrupole peut \^etre calcul\'e avec la loi
newtonienne de la gravitation. C'est la formule ``newtonienne'' du
quadrupole qui permet d'expliquer le ph\'enom\`ene de r\'eaction de
rayonnement dans le pulsar binaire PSR~1913+16, dont on a d\'ej\`a vu
qu'il correspond lui-m\^eme \`a une correction dans l'\'equation du
mouvement \`a l'ordre $(v/c)^5$ soit 2.5PN. La pr\'ecision demand\'ee
pour les binaires spiralantes correspond donc en fait \`a une
contribution d'ordre 3PN+2.5PN c'est-\`a-dire $\sim (v/c)^{11}$ dans les
\'equations du mouvement de la binaire~!

C'est la premi\`ere fois dans l'histoire de la relativit\'e g\'en\'erale
que la r\'ealisation d'exp\'eriences nouvelles suscite des
d\'eveloppements th\'eoriques nouveaux. M\^eme pendant la p\'eriode de
son renouveau, notre vieille dame, confront\'ee \`a des observations et
des tests jamais effectu\'es auparavant, s'\'etait d\'edaigneusement
content\'ee de voir ses pr\'edictions d\'ej\`a ``sur \'etag\`ere''
confirm\'ees. Avant le d\'emarrage de la construction de LIGO et VIRGO
on pensait que les corrections relativistes \`a la formule du quadrupole
d'Einstein n'avaient qu'un int\'er\^et purement acad\'emique (par
exemple ces corrections sont compl\`etement n\'egligeables dans le cas
du pulsar binaire). La pr\'ediction th\'eorique ad\'equate pour les
binaires spiralantes n'existait pas, et il a fallu la d\'evelopper
sp\'ecialement dans le but de fournir les patrons d'ondes n\'ecessaires
\`a l'analyse du signal dans LIGO/VIRGO. Nous avons employ\'e des
m\'ethodes perturbatives analytiques, permettant d'it\'erer les
\'equations d'Einstein sous la forme d'une s\'erie post newtonienne,
d'abord pour des syst\`emes isol\'es g\'en\'eraux, puis dans
l'application \`a des syst\`emes binaires compacts. La pr\'ediction
th\'eorique de la relativit\'e g\'en\'erale, \`a la pr\'ecision 3PN et
m\^eme \`a 3.5PN, est maintenant sur \'etag\`ere (voir la section
suivante). La vieille dame est heureuse.

Un aspect int\'eressant de l'analyse du signal dans VIRGO et LIGO, qui
r\'esulte directement de l'application du d\'eveloppement post
newtonien, est la possibilit\'e d'effectuer des tests nouveaux de la
relativit\'e g\'en\'erale. Dans les patrons d'ondes des binaires
compactes spiralantes, d\'evelopp\'es \`a 3.5PN, existent en effet
plusieurs signatures caract\'eristiques de ce qu'on appelle les
``sillages'' (ou ``tails'') d'ondes gravitationnelles, produits par la
diffusion du rayonnement gravitationnel se propageant dans un
espace-temps distordu par la pr\'esence de la source elle-m\^eme. Cet
effet purement non-lin\'eaire dans la propagation du rayonnement
gravitationnel de sa source vers le d\'etecteur pourra \^etre observ\'e
pour la premi\`ere fois par comparaison des patrons d'ondes avec le
signal r\'eel dans VIRGO/LIGO. On ne conna\^it pas d'autres syst\`emes
que la spirale des binaires compactes, pour lesquels la d\'etection d'un
effet aussi fin que le sillage d'onde gravitationnelle soit possible.

\section{LA FASCINANTE APPROXIMATION 3PN} 

Lors du calcul des patrons d'ondes de binaires compactes spiralantes,
l'approximation $\mathrm{3PN}$, \textit{i.e.} \`a l'ordre $\sim (v/c)^6$
au-del\`a de la formule du quadrupole, s'est av\'er\'ee ``fascinante''
par la complexit\'e des calculs en jeu et la richesse de la th\'eorie
\`a cet ordre. En effet \`a l'ordre $\mathrm{3PN}$ interviennent \`a la
fois les corrections relativistes dans les \'equations du mouvement et
les moments multipolaires de la source, et des effets non-lin\'eaires
associ\'es aux sillages d'ondes gravitationnelles -- une partie de
l'onde se propage \`a une vitesse inf\'erieure \`a $c$ (en moyenne) \`a
cause des diffusions sur la courbure de l'espace-temps induite par la
source. La difficult\'e technique principale est de mettre correctement
en \oe uvre le concept de masse ponctuelle mod\'elisant les corps
compacts. On a recours \`a des m\'ethodes de r\'egularisation du champ
propre de particules ponctuelles.

Le patron d'onde gravitationnelle fournit essentiellement l'\'evolution
temporelle de la phase orbitale $\phi$ de la binaire, par effet de
r\'eaction de rayonnement. La phase est \'ecrite sous la forme d'une
s\'erie post newtonienne, $\phi=\phi_\mathrm{N}\left[1+\sum_n
a_{(n\mathrm{PN})} \,x^n\right]$, o\`u le param\`etre post newtonien
$x\equiv(G m\,\omega/c^3)^{2/3}$ est fonction de la masse totale
$m=m_1+m_2$ et de la fr\'equence orbitale $\omega={\dot\phi}$.
L'approximation newtonienne $\phi_\mathrm{N}$ correspond au ${\dot P}$
du pulsar binaire, \`a ceci pr\`es que pour les binaires spiralantes
l'excentricit\'e de l'orbite est n\'egligeable ($e\simeq 0$). Tous les
coefficients post newtoniens jusqu'\`a l'ordre $\mathrm{3.5PN}$ inclus
sont maintenant connus. \`A titre d'exemple, le coefficient \`a
$\mathrm{3PN}$, qui a \'et\'e le plus difficile \`a calculer, est
donn\'e en fonction du rapport de masse $\nu\equiv m_1m_2/(m_1+m_2)^2$
par l'expression impressionnante
\begin{eqnarray}a_{(\mathrm{3PN})}&=&\frac{12348611926451}{18776862720} 
\nonumber\\ &-& \frac{856}{21}\log\,(16\,x) - \frac{160}{3}\pi^2 -
\frac{1712}{21}C \nonumber\\ &+&
\left(-\frac{15737765635}{12192768}+\frac{2255}{48}\pi^2\right)\nu \nonumber\\
&+& \frac{76055}{6912}\nu^2 -\frac{127825}{5184}\nu^3\,.\nonumber
\end{eqnarray}
Toutes les fractions rationnelles qui apparaissent dans cette formule,
de m\^eme que les nombres irrationnels $\pi$ et la constante d'Euler
$C=\lim_{K\rightarrow\infty}[\sum_{k=1}^{K-1}\frac{1}{k}-\log K]$
(num\'eriquement $C\simeq 0.577$), sont d\'eduites de la relativit\'e
g\'en\'erale et de sa structure non-lin\'eaire. Les expressions comme
celles de $a_{(\mathrm{3PN})}$ sont utilis\'ees pour calculer la
corr\'elation entre le patron d'onde des binaires spiralantes et le
signal de sortie des d\'etecteurs LIGO et VIRGO.

Bien s\^ur il faudrait (id\'ealement) s'assurer que l'approximation
$\mathrm{3PN}$ est proche de la valeur ``exacte''. Cela ne peut pas se
\textit{prouver}, mais n\'eanmoins on peut avoir une id\'ee de la
``convergence'' de la s\'erie post newtonienne en \'etudiant le cas de
l'\'energie du syst\`eme binaire au passage \`a la derni\`ere orbite
circulaire dite ICO. La figure qui suit montre que les ordres post
newtoniens successifs semblent en effet converger vers une valeur
``exacte''. En effet $\mathrm{3PN}$ est tr\`es proche de $\mathrm{2PN}$;
par contre $\mathrm{1PN}$ est clairement insuffisant car l'ICO est une
orbite tr\`es relativiste ($v/c\sim\,0.5$). De plus, la figure montre un
tr\`es bon accord entre $\mathrm{3PN}$ et la valeur calcul\'ee par la
relativit\'e num\'erique dans le cadre d'un mod\`ele approch\'e dit de
``sym\'etrie h\'elico\"idale'' (Gourgoulhon, Grandcl\'ement et
Bonazzola, 2001).
\begin{figure}
\begin{center}
\epsfig{file=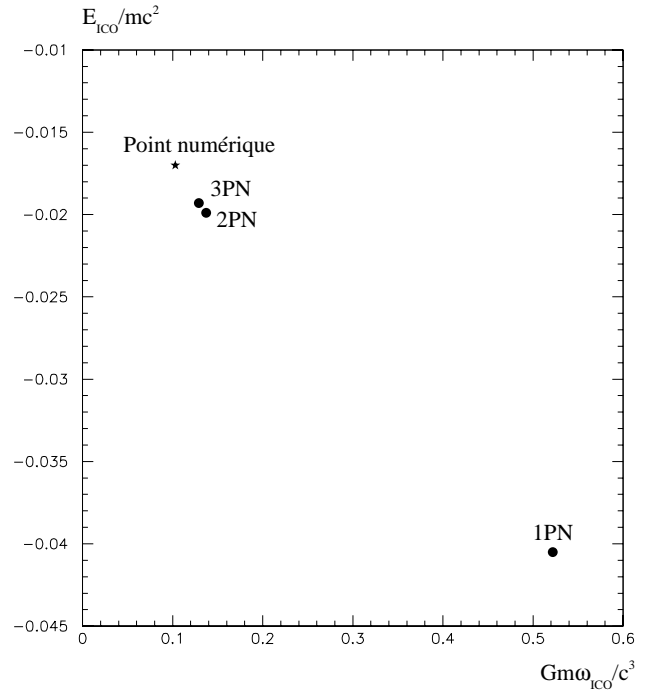,width=3.4in}
\caption{\'Energie d'un syst\`eme binaire de masses \'egales \`a la
derni\`ere orbite circulaire (ICO) en fonction de la fr\'equence
orbitale. On montre les approximations post newtoniennes et le point
calcul\'e par la relativit\'e num\'erique.}
\end{center}
\end{figure}

En outre, on peut faire une estimation de la pr\'ecision du patron
d'onde en comptant le nombre de cycles orbitaux, dans la bande de
fr\'equence de VIRGO, d\^u \`a chacun des ordres post newtoniens. On
trouve qu'\`a l'approximation $\mathrm{3.5PN}$ l'erreur relative sur le
nombre de cycles dans le cas de deux \'etoiles \`a neutrons est
(probablement) inf\'erieure \`a un $1/10\,000^{\text{\`eme}}$.

\bigskip
{\bf Acknowledgements} L'auteur remercie Thibault Damour, Gilles
Esposito-Far\`ese, Guillaume Faye, \'Eric Gourgoulhon et Bala Iyer pour
les nombreuses interactions et/ou collaborations.

 \end{document}